\documentclass{iaus}
\usepackage{graphicx}

\title{Dark Matter Problem in the Local Supercluster}

\author[D. Makarov \and I. Karachentsev]{Dmitry Makarov$^1$ \and Igor Karachentsev$^1$}

\affiliation{$^1$Special Astrophysical Observatory of the Russian Academy of Sciences,
Nizhnij Arkhyz, Zelenchukskij region, Karachai-Cirkassian Republic, Russia 369167 
\break email: dim@sao.ru}

\pubyear{2007}
\volume{244}
\pagerange{447--448}
\jname{Dark Galaxies \& Lost Baryons}
\editors{J. I. Davies \& M. J. Disney, eds.}

\begin{document}

\maketitle

\begin{abstract}
The Local Supercluster is an ideal laboratory to study distribution of luminous and dark matter in the nearby Universe. 
The 1100 small groups have been selected using algorithm based on assumption that a total energy of physical pair of galaxies must be negative. 
The properties of the groups have been considered.
\end{abstract}

\firstsection

\section{Algorithm}
We used an selection algorithm based on assumption that a total energy of physical pair of galaxies must be negative \cite{MK00}. 
%Additional criterion limits a projected distance between galaxies in the pair. 
%The natural bound is the ``zero-velocity'' surface, which separates the collapsing volume against expanding space \cite{S86}.
Individual mass of the galaxies are estimated from their K-luminosity using appropriate value of M/L. 
It is chosen to reproduce the well known nearby groups of galaxies, like the Centaurus~A, M~81, M~83 and IC~342.

\begin{figure}[b]
\includegraphics[width=0.32\textwidth]{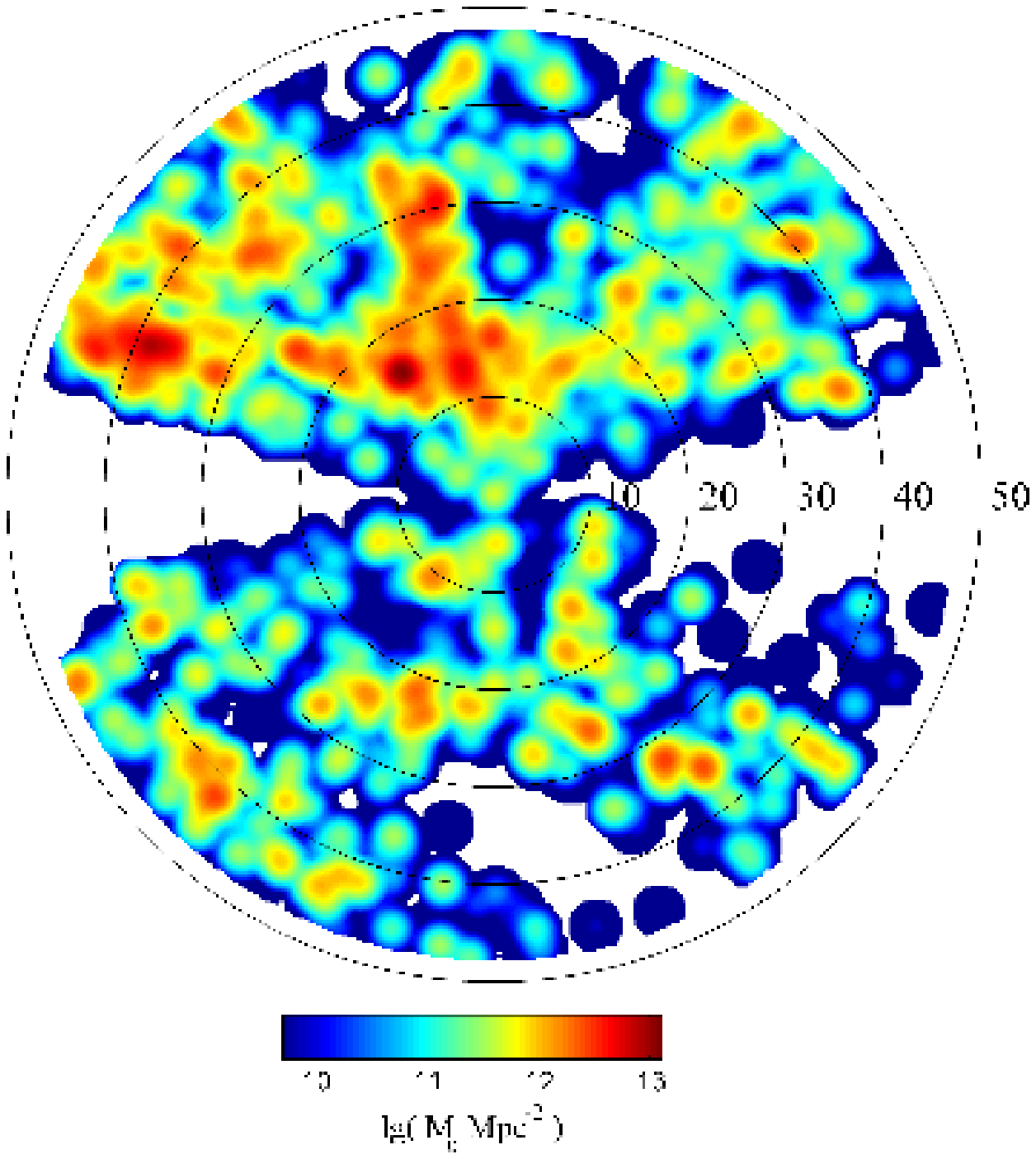}
\includegraphics[width=0.64\textwidth]{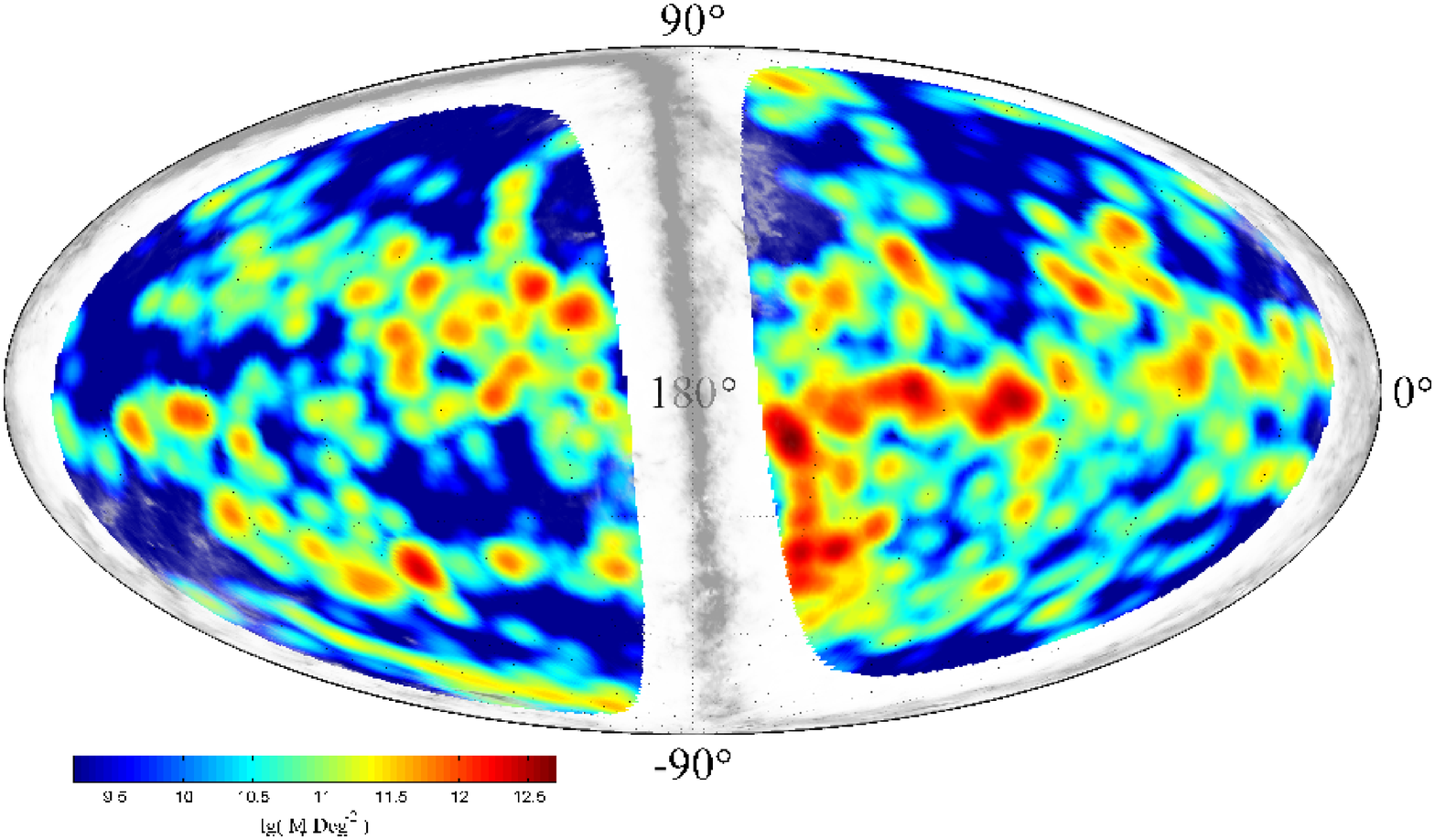}
\caption{The mass distribution in the Local Supercluster. 
The left panel presents distribution of surface mass density in the Local Supercluster plane. 
The right panel shows an all sky map of surface mass density in the Supergalactic coordinates.} 
\label{fig:sky}
\end{figure}

\section{Parameters of the groups}
The 5687 galaxies of 10571 (about 54\%) with $V_{\rm LG}<3500$ and $|b|>15^\circ$ have been gathered in 1127 groups. 
The median velocity dispersion in groups with $\ge5$ members is 88 km~s$^{-1}$, and the median harmonic radius is 222 kpc. 
We find the median mass-to-K-luminosity ratio for the groups to be 26 in solar units using a projected mass estimator \cite{H85}. 
That shows presence of moderate amount of dark matter in the considered groups.

\section{Dark matter in the Local Supercluster}

\begin{figure}[t]
\centerline{\includegraphics[height=0.65\textwidth,angle=90]{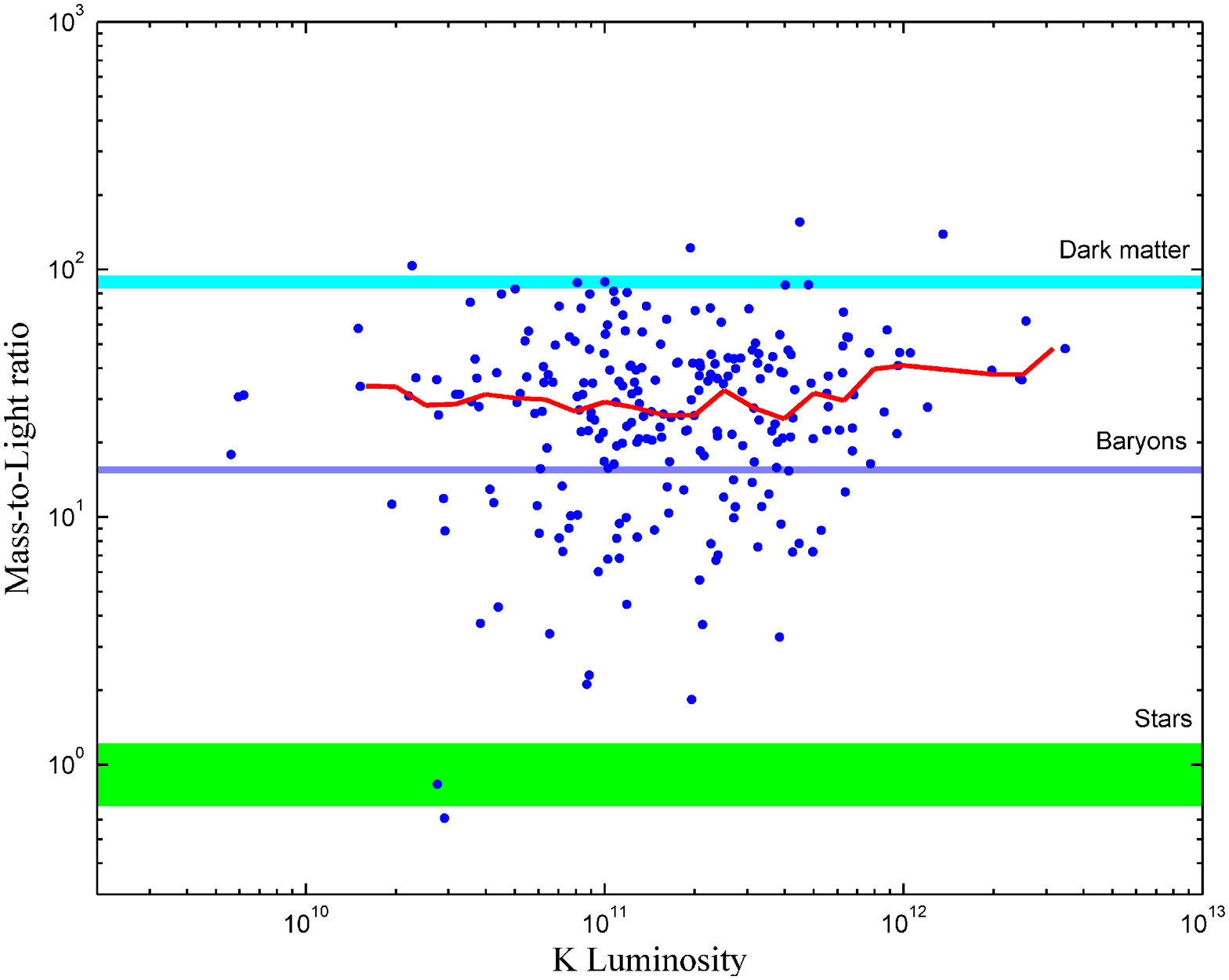}}
\caption{The mass-to-light ratio versus K-band luminosity. 
The dots show the individual groups.
The winding line represents the running median of M/L. 
The horizontal lines show the mean value for stars $M/L=0.95\pm0.27$ \cite{B03}, 
expected M/L ratio for baryonic $\Omega_b h^2=0.02229$ and dark $\Omega_m h^2=0.1277$ 
matter in standard cosmology \cite{S07}.} 
\label{fig:ml}
\end{figure}

Based on our algorithm for group selection we derived the median mass-to-light ratio of groups $M/L_K\sim26$ 
that was used to trace the distribution of mass via light distribution in the Local Supercluster (LSC). 
The mean K-band luminosity density within the considered volume is $j_K=3.96\times10^8$ L$_\odot$ Mpc$^{-3}$, 
being in good agreement with the global luminosity density $4.19\pm0.63\times10^8$ L$_\odot$ Mpc$^{-3}$ from 2dF survey \cite{C01} 
and $4.24\times10^8$ L$_\odot$ Mpc$^{-3}$ from 2MASS and SDSS surveys \cite{B03}. 
Thus, the mean total mass density inside the LSC volume is $1.0\times10^{10}$ M$_\odot$ Mpc$^{-3}$ that corresponds to $\Omega_m=0.07$. 
This value amounts only a quarter of the global density $\Omega_m=0.27$ in standard cosmology \cite{S07}. 
Even the total mass-to-light ratio of the Virgo cluster ($M/L_K\sim62$) is not enough to reach the global value $M/L_K\sim90$ expected for $\Omega_m=0.27$. 
To adjust the ensemble averaged value $\Omega_m=0.07$ with the global density one needs to suppose that a significant part of the dark matter 
on the LSC scale is not associated with system of galaxies.

\begin{acknowledgments}
This work was supported by DFG-RFBR grant 06--02--04017 and RFBR grant 07--02--00005.
\end{acknowledgments}


\begin{thebibliography}{}

\bibitem[(Bell et al. 2003)]{B03}   {Bell et al.} 2003, \textit{ApJ} 149, 289
\bibitem[(Cole et al. 2001)]{C01}   {Cole et al.} 2001, \textit{MNRAS} 326, 255
\bibitem[(Heisler et al. 1985)]{H85}   {Heisler et al.} 1985, \textit{ApJ} 298, 8
\bibitem[(Makarov \& Karachentsev 2000)]{MK00}  {Makarov \& Karachentsev}, 2000, \textit{ASPC} 209, 40
%\bibitem[(Sandage 1986)]{S86}       {Sandage}, 1986, \textit{ApJ} 307, 1
\bibitem[(Spergel et al. 2007)]{S07}    {Spergel et al.} 2007, \textit{ApJS} 170, 377

\end{thebibliography}
\end{document}